\documentclass[num-refs]{nbdt-article}

\usepackage{siunitx}
\usepackage[frozencache=true,cachedir=.]{minted}

\papertype{Original Article}
\paperfield{Education}

\title{Parallel scalable simulations of biological neural networks using TensorFlow: A beginner’s guide}

\abbrevs{CPU, central processing unit; GPU, graphical processing unit; ODE, ordinary differential equation.}

\author{Rishika Mohanta}
\author{Collins Assisi}

\affil[1]{Indian Institute of Science Education and Research, Pune, Maharashtra, India}

\corraddress{Rishika Mohanta and Collins Assisi \newline Indian Institute of Science Education and Research, Pune - 411008, Maharashtra, India}
\corremail{neurorishika@gmail.com, collins@iiserpune.ac.in}

\fundinginfo{Kishore Vaigyanik Protsahan Yojana (KVPY) Fellowship, SB-1712051; DBT–Wellcome India Alliance Intermediate Fellowship, IA/I/11/2500290; IISER Pune}

\runningauthor{Mohanta \& Assisi, 2022}

\begin{document}

\maketitle

\begin{abstract}
Biological neural networks are often modeled as systems of coupled, nonlinear, ordinary or partial differential equations. The number of differential equations used to model a network increases with the size of the network and the level of detail used to model individual neurons and synapses. As one scales up the size of the simulation, it becomes essential to utilize powerful computing platforms. While many tools exist that solve these equations numerically,  they are often platform-specific. Further, there is a high barrier of entry to developing flexible platform-independent general-purpose code that supports hardware acceleration on modern computing architectures such as GPUs/TPUs and Distributed Platforms. TensorFlow is a Python-based open-source package designed for machine learning algorithms. However, it is also a scalable environment for a variety of computations, including solving differential equations using iterative algorithms such as Runge-Kutta methods. In this article and the accompanying tutorials, we present a simple exposition of numerical methods to solve ordinary differential equations using Python and TensorFlow. The tutorials consist of a series of Python notebooks that, over the course of five sessions, will lead novice programmers from writing programs to integrate simple one-dimensional ordinary differential equations using Python to solving a large system (1000's of differential equations) of coupled conductance-based neurons using a highly parallelized and scalable framework. Embedded with the tutorial is a physiologically realistic implementation of a network in the insect olfactory system. This system, consisting of multiple neuron and synapse types, can serve as a template to simulate other networks.

\keywords{TensorFlow, Python, ODE, Jupyter Notebook, Neuronal network, GPU}
\end{abstract}

\section{Motivation}
\begin{figure}[bt]
\centering
\includegraphics[scale=0.7]{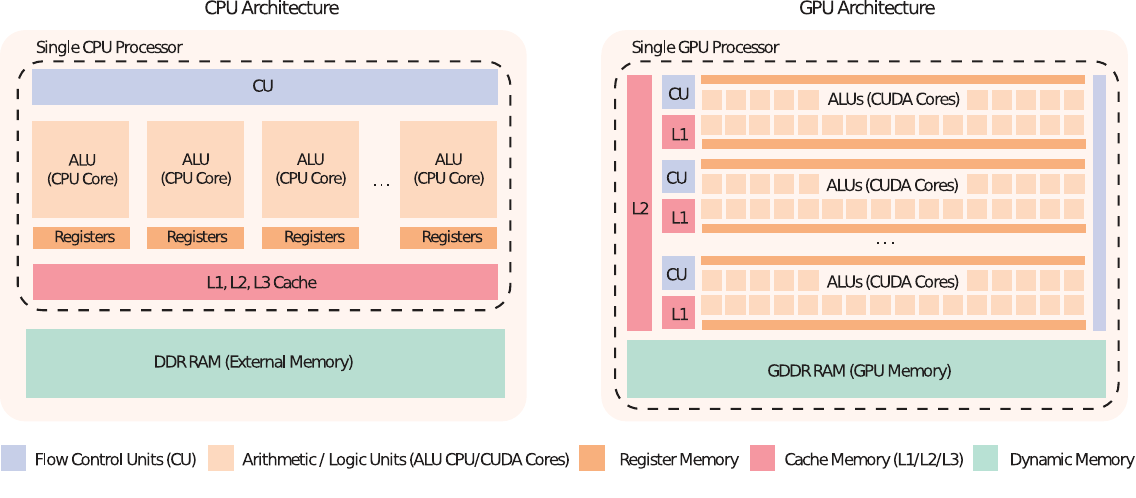}
\caption{\textbf{Comparison between CPU and GPU architecture.}
ALUs (Arithmetic/Logical Units, i.e. CPU Cores/CUDA Cores) are the main circuits that perform all the computations. CPUs (left panel) have fewer ALUs than GPUs (right panel). But, at the same time, the CPU cores can run at a higher clock speed and can do more complex calculations. On the other hand, GPU cores are specialized for linear algebra especially vector operations. Furthermore, a single CPU processor has common memory and control units that must be refreshed every time a new thread of operations is executed. In contrast, GPUs have a hierarchical structure with multiple memory and control units, making them faster at executing multiple threads of operations. This makes GPUs better for simple but highly repetitive tasks executed in parallel, while the CPU is suitable for large complex computations executed serially. However, GPUs have one drawback, external memory (RAM) is typically larger, faster and more robust than internal GPU memory, which is why computation on the GPU is constrained by the amount of data used/generated more often than on CPUs.\cite{Rishika:2019}}
\label{fig:CPUGPU}
\end{figure}

Information processing in the nervous system spans a number of spatial and temporal scales \cite{Churchland1988}. Millisecond fluctuations in ionic concentration at a synapse can cascade into long-term (hours to days) changes in the behavior of an organism. Capturing the temporal scales and the details of the dynamics of the brain is a colossal computational endeavor. The dynamics of single neurons (modeled using one or a few compartments), and small networks of such neurons, can be simulated on a desktop computer with high-level, readable programming languages like Python. However, large networks of conductance-based neurons are often simulated on clusters of CPUs. More recently, graphical processing units (GPUs) have become increasingly available on individual workstations and from cloud services like Google Colab/Cloud and Amazon Web Services (AWS), among others. GPUs typically have hundreds of cores, while CPUs have relatively few (up to 80 at this writing) to execute any computation. Compute-intensive tasks that can be parallelized may be offloaded to the GPU. GPU cores, though many in number, are slower than their CPU counterparts (see Fig.~\ref{fig:CPUGPU} for a comparison of CPUs and GPUs). Therefore, one has to judge whether the simulations can be efficiently completed on a CPU or whether they would benefit from GPU acceleration. Writing code for different platforms is nontrivial and requires a considerable investment of time to master different software tools. For example, implementing parallelism in multi-core shared-memory architectures is often achieved using Open Multi-Processing (OpenMP) with C or C++ \cite{Chapman2007}. Message Passing Interface (MPI) libraries are used to implement code that computes over high-performance computing clusters \cite{Gropp2014}. Compute Unified Device Architecture (CUDA) allows users to run programs on NVIDIA's GPUs \cite{Storti2016}. However, code written for one platform cannot be used on other platforms. This poses a high barrier of entry for Neuroscientists conversant with a high-level programming language, attempting to test simulations on different platforms and scaling up a simulation. To lower this barrier, we created a well-documented tutorial with clear example code to simulate neuronal networks in a platform-independent manner. This tutorial is available in the form of Jupyter notebooks that can be downloaded or run online (\href{https://github.com/neurorishika/PSST}{\texttt{https://github.com/neurorishika/PSST}}) \cite{Rishika:2019}

\begin{figure}[bt]
\centering
\includegraphics[scale=0.35]{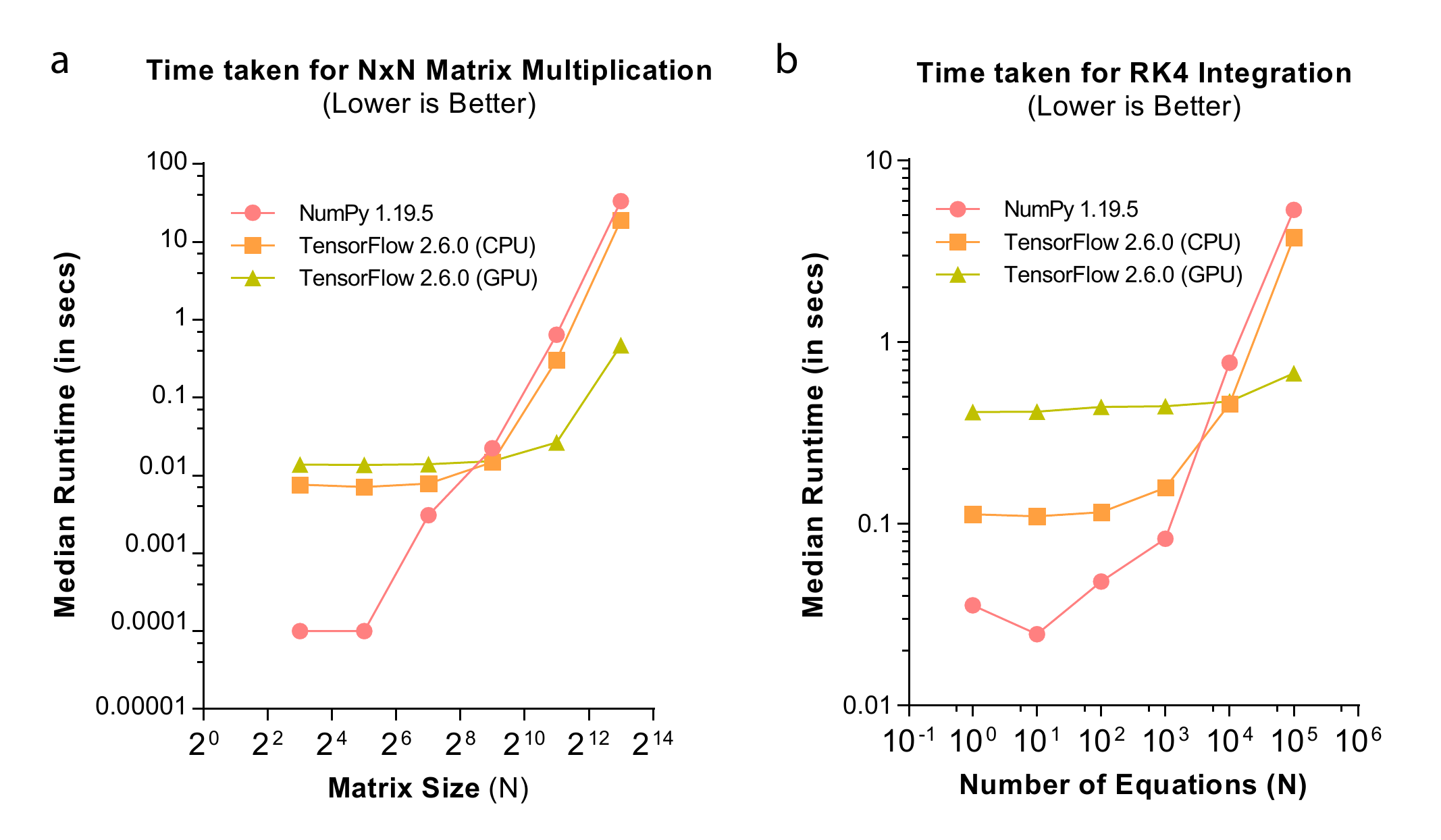}
\caption{\textbf{Comparison of the Matrix Multiplication and Numerical Integration performance of NumPy (CPU only), TensorFlow CPU and TensorFlow GPU.} TensorFlow GPU is much faster in (a) Matrix multiplication of $N\times N$ matrices and (b) RK4-based Numerical Integration of parallel ODEs for large systems and TensorFlow CPU is marginally better than NumPy possibly due to better optimization for parallel computing. Note that here all the equations are independent, but in real systems, they will have interactions for which the computations will be faster on the GPU than the CPU if properly vectorized. Thus, the estimates for the improvement in RK4 integration on TensorFlow GPU are likely underestimated for many practical tasks. Median runtimes were estimated across 20 replicates running on a Google Colab Instance with 1 core, 2 thread 2.30 GHz Intel(R) Xeon(R) CPU and Nvidia a single Tesla K80 GPU with 12 GB GDDR5 Memory.}
\label{fig:comparison}
\end{figure}

Our tutorial leverages TensorFlow, an open-source platform for machine learning~\cite{tensorflow2015-whitepaper} that is highly scalable. Code written using TensorFlow functions can work seamlessly on single cores, multi-core shared memory processors, high-performance computer clusters, and GPUs. We found (as others have ~\cite{tensorflow-api-docs,tfcookbook}), TensorFlow functions can be used to implement numerical methods to solve ODEs. Doing so gave us a significant speed-up even on a single desktop with a multi-core processor compared to similar Python code that did not use TensorFlow functions and operated on a single core (Fig~\ref{fig:comparison}). The code itself was highly readable and could be debugged with ease. Familiarity with the Python programming language and a brief introduction to some TensorFlow functions proved sufficient to write the code. Python is a popular programming language that is used across a number of disciplines and has found a broad user base among Biologists \cite{Ekmekci2016,Bassi2007,primer}. We found that introducing a few TensorFlow functions in Python, an easy addition to a familiar language can bring readers to a point where they can simulate large networks of neurons in a platform-independent manner. Further, by piggybacking on TensorFlow, we could also take advantage of an active TensorFlow developer community and a wide range of Python libraries. 

The tutorials that accompany this paper~\cite{Rishika:2019} were written to address the needs of a group of undergraduate students in our institute. These students came from diverse backgrounds and had a basic introduction to Python during their first semester. Some were interested in working on problems in Computational Neuroscience. Our goal was to introduce them to some of the numerical tools and mathematical models in Neuroscience while also allowing them to tinker with advanced projects. Therefore, we were careful to keep the innards of the code visible \textemdash this included the form of the integrator and the specification of the differential equations. One could argue that our implementation rested upon an extensive software library, namely, TensorFlow, whose workings remained mysterious to a beginning programmer. In coming up with this tutorial, we decided that it was important to get students to engage with actually integrating ODEs and further put them in a position to examine a whole class of problems that require iterative computation in a platform-independent manner. This is not a goal that current simulation environments seek to achieve. We anticipate that the code described here will serve as a starting point to simulate ODEs and could potentially include more sophisticated and faster integration algorithms and methods to manage limitations imposed by memory. We hope that the utility of this tutorial will extend beyond the test cases and the domains we consider here. Towards the end of the tutorial, that many students managed to complete within a day or two, they were in a position to write codes simulating networks of neurons in the antennal lobe  \cite{Bazhenov2001}  (the insect equivalent of the olfactory bulb in mammals), firing rate models of grid cells~\cite{Burak2009}, detailed networks of stellate cells and inhibitory interneurons~\cite{Neru2021} and networks with plastic synapses~\cite{Bazhenov2005}.

\section{How to use the Tutorials}

\begin{figure}[bt]
\centering
\includegraphics[scale=0.5]{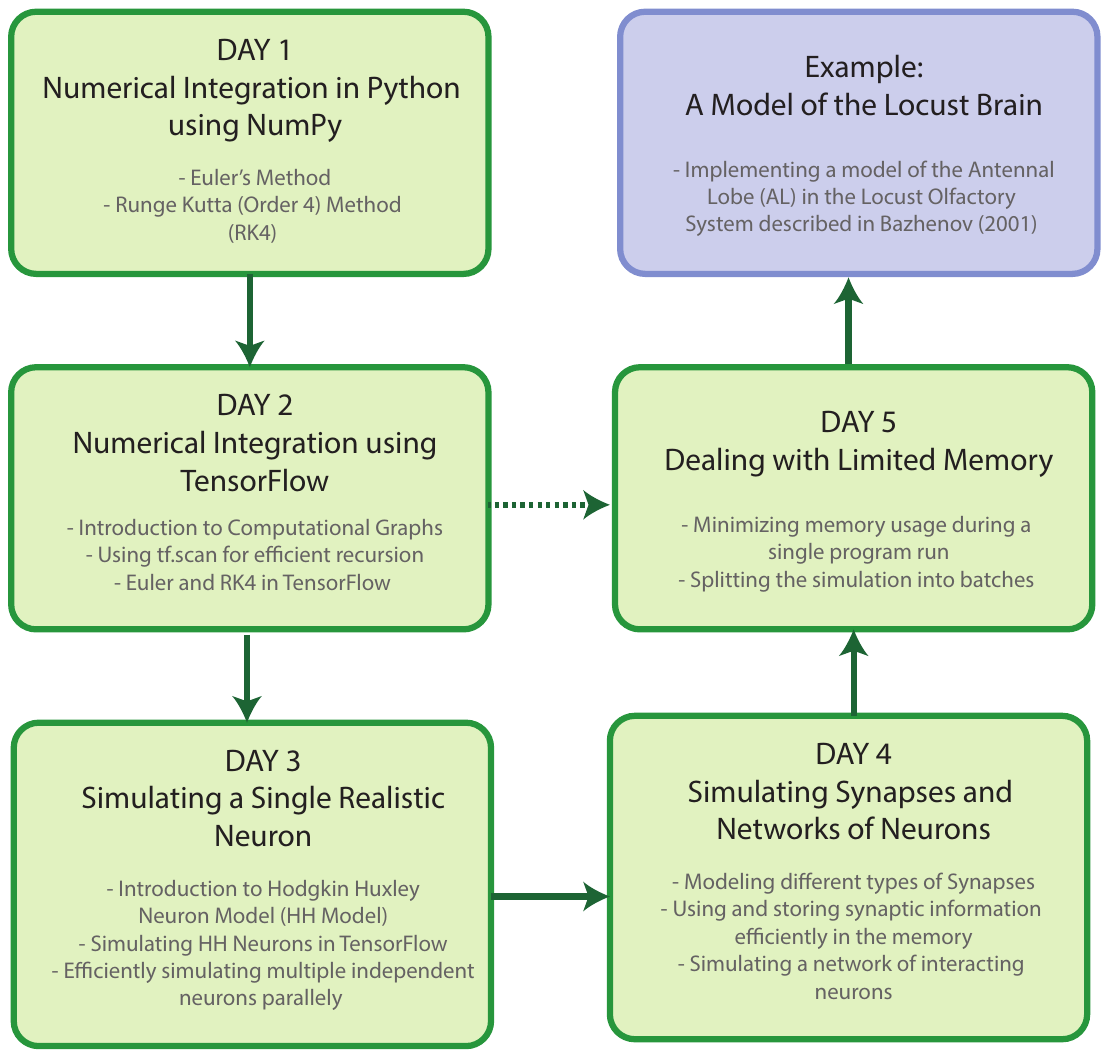}
\caption{Flowchart of the tutorial \cite{Rishika:2019} accompanying this paper}
\label{fig:flowchart}
\end{figure}

This paper and the accompanying tutorials may be used in a classroom setting to supplement the material taught in a  Computational Neuroscience or Mathematical Biology course, where integrating ordinary differential equations in different contexts would be a core learning objective. We believe that the tutorial may also be appropriate for self-study by students who are familiar with the Python programming language and popular libraries such as NumPy and Matplotlib that are used to represent some of the data structures and plot the results. These libraries are well documented with excellent introductory guides. Fig.~\ref{fig:flowchart} provides an overview of the tutorials.  Readers who are interested in solving differential equations in other domains will find the tutorial on Days 1, 2, and 5 self-contained. On days 3 and 4, we introduce conductance based neurons using the Hodgkin-Huxley formalism. Our introduction to the Hodgkin-Huxley model is terse and will benefit from supplementary classroom instruction or from self-study using one of several excellent guides on the topic (\cite{gerstnerMOOC,Johnston1995,Dayan2005}). The tutorials are linked in the Supporting information and are available as Jupyter notebooks (.ipynb files). The notebooks can be viewed or run online using Binder, Google Colab or Kaggle Notebook. The respective links are available in each notebook. An online version of the tutorials is also available as a JupyterBook linked in the supporting information. Please enable GPU usage on Google Colab and Kaggle Notebooks for the best performance. We also provide .html files that can be read using any browser. To run the notebooks locally, we recommend that readers install Python 3.6 or above, Jupyter Notebook, NumPy 1.20.or above~\cite{numpy}, Matplotlib 3.4 or above~\cite{matplotlib}, and TensorFlow 2.8 or above using the Anaconda distribution of Python 3.9. We suggest that users utilize TensorFlow 2.x with eager execution disabled for optimal stability and performance.

\section{Computational graphs and TensorFlow}

\begin{figure}
\centering
\includegraphics[scale=0.8]{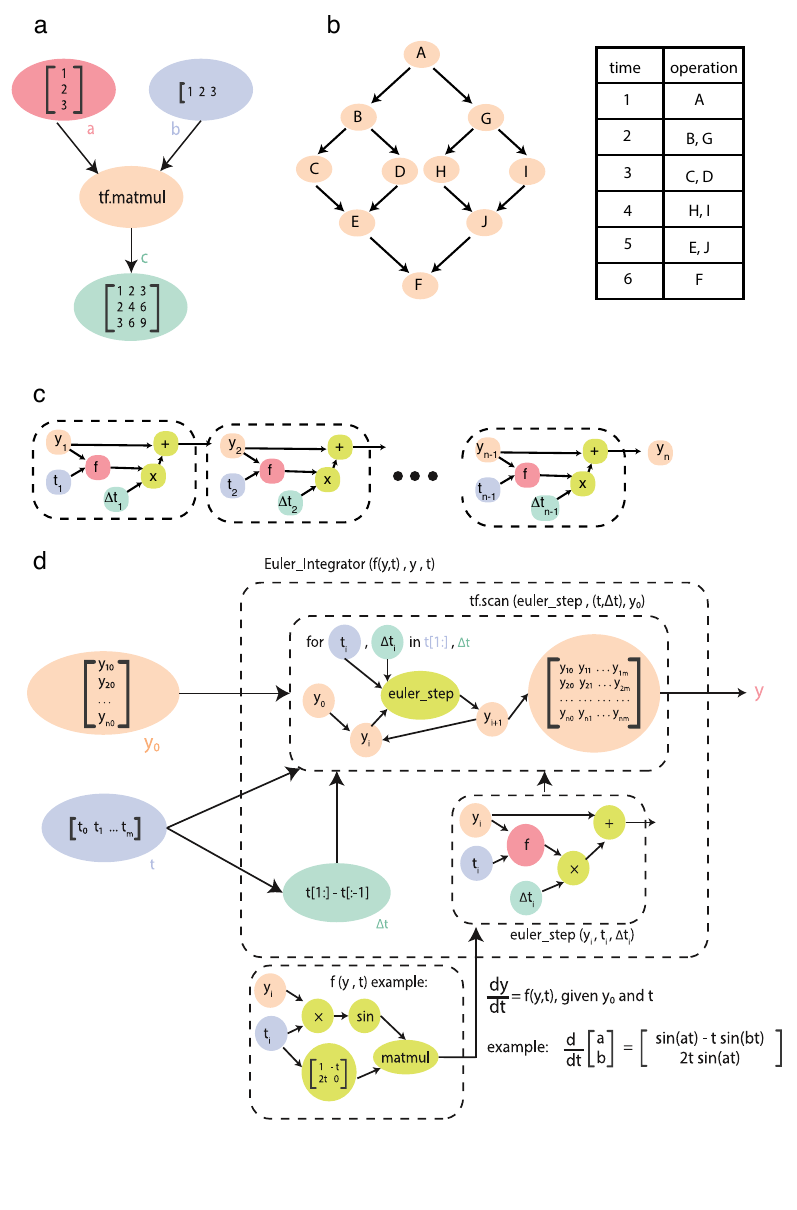}
\caption{\textbf{Computational graphs for TensorFlow.} a. Example of a simple computational graph to multiply two vectors $a$ and $b$ and store it in $c$. b. A  simple computational graph is shown on the left. An efficient schedule (right) to compute the operations assuming each operation takes the same time and two processors are available. The table on the right shows the time steps at which specific operations are completed. c. A for loop implemented as a computational graph. d. A schematic showing the different steps involved in implementing the Euler integrator as a computational graph to solve an initial value problem of the form $\frac{dy}{dt}=f(y,t)$ given an initial condition $y_{0}$ (left top), a time vector $t$ (left bottom) and the function $f(y,t)$ (bottom). The graph for $f(y,t)$ is fed to the \texttt{euler\_step} function which is in turn fed to the \texttt{tf.scan} function along with the time-step vector $\Delta t$ and the initial condition $y_{0}$. \texttt{tf.scan} iterated recursively over the time difference and calculates the integrated time-evolved output $y$.}
\label{fig:compGraph}
\end{figure}

TensorFlow is an open-source library developed by researchers and engineers in the Google Brain team. TensorFlow has a number of functions that make it particularly suitable for machine learning applications. However, it is primarily an interface for numerical computation~\cite {tensorflow2015-whitepaper}. All computations in TensorFlow are specified as \textit{computational graphs}. Computational graphs, also known as data flow graphs, are directed graphs (nodes connected by arrows) where the nodes represent operations (for example, addition or multiplication), while incoming edges to each node represent the data stored as tensors (scalars, vectors, matrices, and higher-dimensional arrays) - these are the actual values that are operated upon. The output of the computation is also a tensor. For example, consider the following computation where two vectors $a$ and $b$ serve as inputs to the node, a matrix multiplication operation, that produces a matrix $c$ as output (Fig~\ref{fig:compGraph}a).

The following program implements the computation described in Fig~\ref{fig:compGraph}a.

\begin{minted}[linenos]{python}
# Creating nodes in the computation graph
a = tf.constant([[1.],[2.],[3.]], dtype=tf.float64) # a 3x1 column matrix
b = tf.constant([[1.,2.,3.]], dtype=tf.float64) # a 1x3 row matrix
c = tf.matmul(a, b)
# To run the graph, we need to create a session.
# Creating the session initializes the computational device.
with tf.Session() as sess:
   output = sess.run(c)
print(output)
\end{minted}

Any computation may be thus defined as a sequence of TensorFlow operations acting on tensors. The final output is computed by passing the data through the directed edges of the computational graph. By specifying the computational graph, we also specify the dependencies between operations. One can thus split the graph into smaller chunks or sub-graphs that can be independently computed by different devices that coordinate with each other. Consider, for example, the computational graph shown in Fig~\ref{fig:compGraph}b. Here, each node represents an operation. Assume that the time taken to complete each operation is the same, and that we have two processors available to complete the computation. As the computation progresses, some operations  may be scheduled in parallel. The computational graph states the dependency between operations, (operations E and J can only be completed once C and D, and H and I are complete). This allows us to efficiently schedule operations such that the entire program can be completed in the least number of time units (in Fig~\ref{fig:compGraph}b, the number of operations is 6 \footnote{Example based on lecture notes of a course on Functional Programing by David Walker, Princeton University, https://www.cs.princeton.edu/~dpw/courses/cos326-12/notes/parallel-schedules.php}). Further, common sub-graphs, if any, may be eliminated to speed up the computation. Computational graphs make it possible to develop programs that are fast, device-independent, and scalable across CPUs, GPUs, and clusters of servers. These graphs are not unique to TensorFlow. Other machine learning frameworks such as PyTorch and Theano also utilize a similar system in the background. Therefore, the concepts developed in this tutorial can also be translated to other frameworks.

\section{Brain simulations as computational graphs}
Days 1 and 2 of the tutorial develop iterative methods to integrate ordinary differential equations (ODEs). In particular, we present simple implementations of Runge-Kutta methods of order one and four and elaborate on some of the peculiarities associated with implementing these recursive methods using TensorFlow. An example of ODEs at the core of our tutorial are the Hodgkin-Huxley equations that describe the dynamics of action potential generation and propagation in the giant axon of the squid~\cite{Huxley1952}. The Hodgkin-Huxley equations have kindled a revolution in our understanding of brain function, touching multiple scales of organization, ranging from the dynamics of ion channels to the collective behavior of networks of neurons (see \cite{Catterall14064} for a review). Alan Hodgkin and Andrew Huxley arrived at these equations through a series of clever experiments that tested the limits of the technology available at the time. It also tested the limits of computational tools available. In order to compute action potentials, Huxley numerically integrated the equations using a hand-operated Bunsviga mechanical calculator. The calculation took nearly three weeks to complete~\cite{Hodgkin1976}. On day 3 of the tutorial, we use a fourth-order Runge-Kutta integrator (implemented on day 2 of the tutorial) to integrate the Hodgkin-Huxley equations. Further, since much of our implementation is geared towards simulating networks of neurons, the tutorial dwells on conductance-based neuronal models and provides specific pointers on how to integrate the differential equations describing neuronal networks efficiently (days 3 and 4). Consider an initial value problem of the form,

\begin{equation}
\frac{dy}{dt} = f(y, t) \hspace{0.3in} \text{with} \hspace{0.3in} y(t_{0}) = y_{0}
\label{eq:ode}
\end{equation}
where, $y$ is an $N-$dimensional vector and $t$ typically stands for time. The function $f(y,t)$ may be a nonlinear function of $y$ that explicitly depends on $t$.  Euler's method iteratively solves eqn~(\ref{eq:ode}). Here we start from $y(t=t_{0})=y_{0}$ and compute the solution at subsequent time points ($t_{0}+\Delta t,t_{0}+2\Delta t,t_{0}+3\Delta t \dots $). The solution at each step can be derived by truncating a Taylor series after the first term. That is, the solution at time $t_{0}+\Delta t$ is given by,

\begin{equation}
y(t_{0}+\Delta t) = y_{0} + \Delta t\frac{dy}{dt} + \mathcal{O}(\Delta t^2)
\label{eq:euler}
\end{equation}

where $\frac{dy}{dt}=f(y,t)$. The higher order terms $\mathcal{O}(\Delta t^2)$ are ignored in this approximation.

Euler integration is essentially a recursive process over time-series $[t_{0},t_{1},t_{2},\dots t_{n}]$. The eqn~(\ref{eq:euler}) thus can be written as a recursive function $F$ such that $X_{i+1}=F(X_i,t_i,\epsilon_i)$. The final solution of the ODE becomes\\ $[X_0,F(X_0,t_0,\epsilon_0),F(F(X_0,t_0,\epsilon_0),t_1,\epsilon_1)...]$. In Python, this recursive function can be implemented using a "for" loop. However, implementing a similar loop in TensorFlow would result in a computational graph consisting of a long chain of sub-graphs,  each sub-graph representing a single iteration of the loop  (Fig~\ref{fig:compGraph}c ). Note, each box (demarcated by the dashed lines) represents a single step of a first order Euler integrator. The integrator itself is detailed in Fig~\ref{fig:compGraph}d. The size of this computational graph increases with the number of iterations and results in a memory-intensive way to compute a recursive function. TensorFlow provides a more efficient solution in the form of the function \texttt{tf.scan} ~\cite{tensorflow-api-docs} to iterate over the time series. \texttt{tf.scan} takes in 3 inputs: (i) a recursive function (ii) the list to iterate over and (iii) the initial value. If the initial value is not specified, it uses the first element of the list as an initial value. For example, consider the following program that calculates the cumulative sum over a list. Every step involves adding an element from the list onto the last addition.

\begin{minted}[linenos]{python}
# define the recursive function that takes the accumulated
# value and the additional input from a list.
def recursive_addition(sum_till_now,next_value):
    return sum_till_now + next_value
# define the list over which we iterate
elems = np.array([1, 2, 3, 4, 5, 6])
# accumulate with the starting number 5
cum_sum = tf.scan(recursive_addition, elems, tf.constant(5))
with tf.Session() as sess:
    output = sess.run(cum_sum)
print(output)
# This prints :
#[ 6  8 11 15 20 26]
\end{minted}

We can use the same principle to create a recursive function that calculates each step in Euler's method (see accompanying tutorial \cite{Rishika:2019} for the implementation). The computational graph in Fig \ref{fig:compGraph}c and d is a schematic showing the data flow in Euler's method. Here the function is called recursively to compute and accumulate the value of the state variable $[\begin{matrix} y_{i1} & y_{i2} & \dots & y_{iN} \end{matrix}]^{T}$ at time $t_{i}$. TensorFlow includes a visualization tool, TensorBoard, that may be used to create a data flow graph. The graph visualized in Fig \ref{fig:compGraph}d was not generated using this tool but is a schematic that outlines the steps involved in computing the solution of (\ref{eq:ode}) using Euler's method. In addition to Euler's method, the tutorial (\cite{Rishika:2019}) implements the Runge-Kutta method of order 4 (abbreviated as RK4). While the error per iteration is $\mathcal{O}(\Delta t^5)$ in the RK4 method, it requires additional computations to calculate the value of the solution at each time step. 

Each iteration operates upon multiple tensors and computes a solution. TensorFlow has several built-in functions that speed up Tensor computations using available multi-core CPUs and GPU hardware. Two components of the code where a significant speed-up can be achieved are,

\begin{enumerate}
\item \textbf{Iterations of Numerical Integration:} TensorFlow can be set to implement numerical integration algorithms (Euler/RK4) as computational graphs that are compiled and run efficiently (see Day 1 of the accompanying tutorial \cite{Rishika:2019}).
\item \textbf{Vectorized Functional Evaluations:} Converting loops into array operations is often termed `vectorization'. Array operations are computed by highly optimized functions and are, as a result, nearly an order of magnitude faster to evaluate. The form of the equations that describe the neural dynamics are similar across neurons even though the specific parameters may vary. A large number of such equations may thus be vectorized, eliminating lengthy for loops (see Day 3 of the accompanying tutorial \cite{Rishika:2019}).

Say $\vec{X}=[V,m,n,h]$ is the state vector of a single neuron and its dynamics are defined using parameters $C_m,g_K,...E_L$. The equations governing the dynamics of the system is of the form: 

\begin{eqnarray}\frac{d\vec{X}}{dt} = [f_1(\vec{X},C_m,g_K,...E_L),f_2(\vec{X},C_m,g_K,...E_L)...f_m(\vec{X},C_m,g_K,...E_L)]\end{eqnarray}

We can convert these equations to a form in which some evaluations are done as array operations and not as for loops. Despite the parameters being different, the functional forms of the equations are similar for the same state variable of different neurons. Thus, the trick is to reorganize $\mathbf{X}$ as \\ $\mathbf{X'}=[(V_1,V_2,...V_N),(m_1,m_2,...m_N),(h_1,h_2,...h_N),(n_1,n_2,...n_N)]=[\vec{V},\vec{m},\vec{h},\vec{n}]$. And the parameters as $[\vec{C_m},\vec{g_K}] = [C_{m_{1}}\dots C_{m_{N}},g_{K_{1}}\dots g_{K_{N}}]$ and so on.

The advantage of this re-ordering is that the system of differential equations of the form,
\begin{eqnarray}\frac{dV_i}{dt}=f(V_i,m_i,h_i,n_i,C_{m_i},g_{K_i}...)\end{eqnarray}

where $i = 1, ...,N$, can be rewritten in vector form as,

\begin{eqnarray}\frac{d\vec{V}}{dt}=f(\vec{V},\vec{m},\vec{h},\vec{n},\vec{C_m},\vec{g_K}...)\end{eqnarray}
The `vectorized' equations now read,
\begin{eqnarray}\frac{d\mathbf{X'}}{dt}= \Big[\frac{d\vec{V}}{dt},\frac{d\vec{m}}{dt},\frac{d\vec{h}}{dt},\frac{d\vec{n}}{dt}\Big]\end{eqnarray}
\end{enumerate}
These equations may then be solved efficiently using highly optimized, in-built array operations.

\section{Circumventing memory constraints}

Modern architectures such as GPUs/TPUs allow several parallel computations. However, GPU/TPUs are also subject to certain constraints that need to be circumvented to use them effectively. One such constraint is that of memory. In order to allow extensive parallelization, GPUs/TPUs are equipped with high bandwidth but smaller memory than traditional CPUs, a trade-off between memory size and speed due to cost constraints (Fig. \ref{fig:comparison}). Further, as the GPU is managed by a driver and not directly by the operating system, it also faces memory scheduling constraints that reduce the effectiveness of dynamic memory allocation.

Using Python and TensorFlow allows us to write code that is readable, parallelizable, and scalable across a variety of computational devices. However, given the potential scale of our problem (integrating large networks of neurons), our implementation is memory intensive. The iterators in TensorFlow do not follow the normal process of memory allocation and garbage collection. Since TensorFlow is designed to work on diverse hardware like GPUs and distributed platforms, memory allocation is done adaptively during the TensorFlow session and not cleared until the Python kernel has stopped execution. To deal with these issues, we need to carefully optimize how we build our computational graphs in different ways:

\begin{enumerate}
\item \textbf{Optimize graphs to minimize memory required:} In a network with $n$ neurons, there are at most $n^2$ synapses of each type. The actual number of synapses may be much smaller. Arranging synaptic variables as an $n\times n$ matrix is computationally efficient as it reduces the computation at each step to a matrix multiplication that is optimized in TensorFlow. However, storing an $n\times n$ matrix is memory intensive. In many neural networks, where synaptic connections are few in number, several of these $n^2$ elements will be set to zero. Therefore, we store only the non-zero elements of the $n \times n$ matrix, also termed a sparse representation. This is efficient when updating synaptic variables do not require array operations. However, when array operations, such as matrix multiplications, become necessary in order to eliminate time-consuming for loops, we switch from a sparse to a dense matrix representation (Fig. \ref{fig:sparse}). This makes the implementation of the RK4 integrator memory efficient while also minimizing computation time. (see Day 4 of the tutorial for details \cite{Rishika:2019})

\begin{figure}[H]
\centering
\includegraphics[scale=0.8]{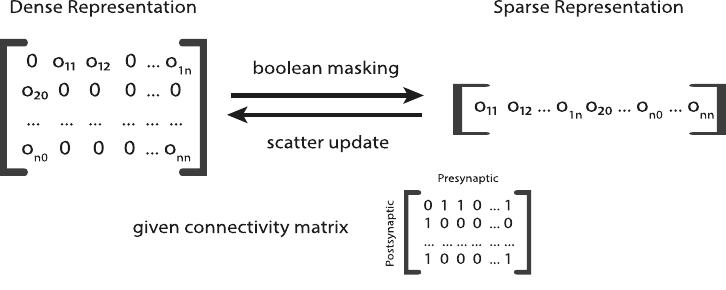}
\caption{\textbf{Dense and sparse representations of Synaptic variables.} Often, when looking at realistic network of neurons, not all neurons are connected to each other, i.e. every neuron forms a synapse with a small fraction of other neurons. Thus, the values of any synaptic variables can be represented as either a dense or sparse variable both of which offer different benefits. Consider the example of a network with $n$ neurons. A dense representation will have $n^2$ values many of which will be 0, while it will take a larger amount of memory to store it, it can be used to write synaptic updates as matrix equations which can be sped up using TensorFlow. However, since memory is a limited resource for computations on a GPU, when the dense representation is not needed, it becomes more efficient to store the synaptic variables as a sparse representation which will consume only as much memory as the number of synapses.}
\label{fig:sparse}
\end{figure}

\item \textbf{Split execution into batches:} The maximum memory used by the computational graph is approximately two times the size of the solution matrix when the computation finishes and copies the final data into the memory. Larger network sizes and longer simulation times result in larger solution matrices. For any given network, the maximum simulation length is limited by the memory available. One way to increase the maximum length is to divide the simulation into smaller batches. The queuing time between batches adds some overhead that slows down our code but allows longer simulation times. Further, memory is cleared only when the Python kernel is closed. By calling a Python script iteratively to execute successive batches of the simulation, we can achieve indefinitely long simulations while optimizing the usage of the available memory.

\item \textbf{Disabling eager execution}: Earlier versions of TensorFlow used a static computational graph that had to be compiled before the simulation was run. This allowed TensorFlow to determine redundancies in the graph and optimize scheduling various computations assigned to different cores. TensorFlow v2.0 introduced \textit{eager execution} which allows dynamic memory allocation and access without explicitly having to create and compile computational graphs in a session. Eager execution is enabled by default in TensorFlow v 2.0. While this is more intuitive and user-friendly than defining computational graphs, we found that it predictably slowed down our simulations. We were able to circumvent this issue by disabling eager execution in TensorFlow v2.0.

\end{enumerate}

\section{Conclusion}
In this paper and accompanying tutorial, our attempt was to delve into the practical question, how does one simulate a network of neurons in a platform-independent manner? There are several simulation packages that simulate brain dynamics at various scales ranging from the sub-cellular ~\cite{10.3389/neuro.11.006.2008, carnevale_hines_2006} to cellular ~\cite{10.3389/neuro.11.006.2008, carnevale_hines_2006, Bower1998, Gewaltig:NEST, Stimberg2019,10.3389/neuro.01.026.2009} and inter-areal networks \cite{10.3389/fninf.2013.00010}. These toolboxes perform with remarkable efficiency and employ sophisticated user interfaces. However, switching platforms, when possible, requires re-writing some components of the code and incorporating different libraries \cite{Kumbhar2019,Stimberg2020}. The degree of difficulty in achieving this can vary across toolboxes. At this point, it is clear that horizontal scalability remains a problem that will require significant changes to the code written for single and multi-core CPUs to be deployed on multi-node HPCs and GPUs. Simulation packages provide sophisticated user interfaces that hide the actual computations (numerical integrators and support functions) to allow the end-user to focus on the scientific question at hand. As the suite of features implemented in different simulation environments increases, it becomes difficult to grasp the details of the implementation comprehensively. Our approach, in contrast, was to minimize the implementation to only those elements that were necessary to simulate a network of neurons efficiently. In doing so, we found that the entirety of the code became accessible to novice programmers who could progress from understanding the basics of numerical integration to actually implementing large scale simulations in a platform-independent manner using TensorFlow.

In recent years, there have been several advances in the hardware available for large-scale computing. In 2016, Google announced a new hardware architecture known as TPUs (Tensor Processing Units) that were specifically designed for use with the TensorFlow library. TPUs utilize a specialized Matrix Multiplication Unit that maximizes parallel computation and are capable of much faster matrix multiplication than CPUs or GPUs. As a result, TPUs can further speed up the the integration of ODEs and other numerical simulations. While, the code we presented in this tutorial can be run on a system with available TPUs, the allocation of tasks to different TPU cores is not currently automated and must be done manually. And so, using TPUs efficiently may require a few additional steps which are outside the scope of this tutorial. However, as TPUs are becoming more popular, it is likely that future versions of the TensorFlow library will be able to automatically allocate tasks to different TPU cores.

\section*{supporting information}
The code in this tutorial is implemented as a series of Jupyter notebooks that can be found in the following GitHub repository: \href{https://github.com/neurorishika/PSST}{\texttt{https://github.com/neurorishika/PSST}} or OSF archive \href{http://doi.org/10.17605/OSF.IO/YBZKQ }{\texttt{http://doi.org/10.17605/OSF.IO/YBZKQ }}. It is also available as an online book at the following link: \href{https://neurorishika.github.io/PSST}{\texttt{https://neurorishika.github.io/PSST}}.

\section*{acknowledgments}
RM received a KVPY fellowship SB-1712051 and support from IISER Pune. CA was funded by DBT–Wellcome India Alliance through an Intermediate fellowship IA/I/11/2500290 and IISER Pune. We thank members of the Assisi and Nadkarni labs at IISER Pune and several students who tested the code. We thank Prof. Maxim Bazhenov for discussions and code related to the insect antennal lobe model. We also acknowledge the National Supercomputing Mission (NSM) for providing computing resources of ‘PARAM Brahma’ at IISER Pune, which is implemented by C-DAC and supported by the Ministry of Electronics and Information Technology (MeitY) and Department of Science and Technology (DST), Government of India.

\section*{conflict of interest}
Authors declare no competing interests.
\bibliography{nbdt_collins_rishika}
\end{document}